\documentclass[letterpaper]{jpconf}
\usepackage{graphicx}
\newcommand{\gamgam}{\gamma\gamma}
\newcommand{\jpipi}{\pi^{+}\pi^{-}J/\psi}

\newcommand{\deltm}{\mathrm{M}(\pi^{+}\pi^{-}l^{+}l^{-}) - 
		    \mathrm{M}(l^{+}l^{-})}

\newcommand{\jll}{J/\psi \rightarrow l^{+}l^{-}}
\newcommand{\jee}{J/\psi \rightarrow e^{+}e^{-}}
\newcommand{\jmm}{J/\psi \rightarrow \mu^{+}\mu^{-}}

\newcommand{\Gammagg}{\Gamma_{\gamgam}(X(3872))}
\newcommand{\Gammaee}{\Gamma_{ee}(X(3872))}

\newcommand{\xBR}{{\cal{B}}(X\rightarrow\jpipi)}

\newcommand{\xGGUL}{(2J+1)\Gammagg\xBR}
\newcommand{\xISRUL}{\Gammaee\xBR}

\begin{document}
\title{Search for $X$(3872) in $\gamgam$ Fusion and ISR at CLEO}

\author{Peter Zweber (CLEO Collaboration)}

\address{Northwestern University, Evanston, Illinois 60208, USA}

\ead{petez@mail.lns.cornell.edu}

\begin{abstract}%less than 200 words
We report on a search for the $X$(3872) state using 
15.1 fb$^{-1}$ of $e^{+}e^{-}$ annihilation data taken with the CLEO III 
detector in the $\sqrt{s}$ = 9.46-11.30 GeV region.  Separate searches for the
production of $X$(3872)  in untagged $\gamgam$ fusion and $e^{+}e^{-}$ 
annihilation following initial state radiation (ISR) 
are made by taking advantage of the unique correlation of $\jll$
in $X$(3872) decay into $\jpipi$.  No signals are observed in either case, 
and 90$\%$ confidence upper limits are established as 
$\xGGUL$ $<$ 12.9 eV and $\xISRUL$ $<$ 8.3 eV.
\end{abstract}

\section{Introduction}
The Belle Collaboration reported the observation of a narrow 
state, $X$(3872), in the decay $B^{\pm}$ $\rightarrow$ $K^{\pm}X$, $X$ 
$\rightarrow$ $\jpipi$, $\jll$ ($l$ = $e,\mu$) \cite{xBELLE}.  
The observation was confirmed by the CDF II \cite{xCDFII}, D{\O} \cite{xD0},
and {\slshape{B{\scriptsize{A}}B{\scriptsize{AR}}}} \cite{xBABAR} 
Collaborations with consistent results, i.e., M($X$) = 3872.0 $\pm$ 1.4 
MeV/c$^{2}$ and $\Gamma(X)$ $\le$ 3 MeV/c$^{2}$.

Many different theoretical interpretations of the nature of the 
$X$(3872) state and its possible quantum numbers have been 
proposed \cite{xccbarbound,xmolecule,xvectorglueball,xother}.  These 
include that (a) $X$(3872) is a charmonium state \cite{xccbarbound}; 
(b) $X$(3872) is a $D^{0}\bar{D}$$^{*0}$ ``molecular'' state \cite{xmolecule}; 
and (c) $X$(3872) is an exotic state \cite{xvectorglueball}. 

No positive signals for $X$(3872) have been observed in searches for the 
decays $X$(3872) $\rightarrow$ $\gamma\chi_{c1}$ \cite{xBELLE}, 
$\gamma\chi_{c2}$, $\gamma J/\psi$, 
$\pi^{0} \pi^{0} J/\psi$ \cite{xOmegaJpsiBELLE}, 
$\eta J/\psi$ \cite{xJpsiEtaBABAR}, $D^{+}D^{-}$, $D^{0}\bar{D}$$^{0}$, 
and $D^{0}\bar{D}$$^{0}\pi^{0}$ \cite{xDDBELLE}, or for possible charged 
partners of $X$(3872) \cite{xChargedBABAR}.  Yuan, Mo, and 
Wang \cite{xISRBES} have used 22.3 pb$^{-1}$ of BES data at $\sqrt{s}$ = 
4.03 GeV to determine the 90$\%$ confidence upper limit of 
$\xISRUL$ $<$ 10 eV for ISR production of $X$(3872).  
Belle \cite{xOmegaJpsiBELLE} has recently reported a small enhancement in 
the $\pi^{+}\pi^{-}\pi^{0}J/\psi$ effective mass near the $X$(3872) mass.  

The variety of possibilities for the structure of $X$(3872) suggests that 
it is useful to limit the $J^{PC}$ of $X$(3872) as much as possible.  The 
present investigation is designated to provide experimental constraints for 
the $J^{PC}$ of $X$(3872) by studying its production in $\gamma\gamma$ 
fusion and ISR, and its decay into $\pi^{+}\pi^{-}J/\psi$ \cite{xCLEO}.  
Production of $X$(3872) in $\gamma\gamma$ fusion can shed light on the 
positive charge parity candidate states, charmonium states 2$^{3}$P$_{0}$, 
2$^{3}$P$_{2}$ and 1$^{1}$D$_{2}$ \cite{xccbarbound}, and the $0^{-+}$ 
molecular state \cite{xmolecule}.  ISR production can address the $1^{--}$ 
vector state.
 
\section{Event Selection}
The data consist of a 15.1 fb$^{-1}$ sample of $e^{+}e^{-}$ collisions at or 
near the energies of the $\Upsilon(nS)$ resonances ($n=1$--$5$) and in the 
vicinity of the $\Lambda_{b}\bar{\Lambda}_{b}$ threshold 
collected with the CLEO III detector \cite{CLEOIIIDetector}.  
Table \ref{tab:infotable} lists the six different initial center-of-mass 
energies and $e^{+}e^{-}$ integrated luminosities at each.  

\begin{table}
\caption{\label{tab:infotable}Data sample for the present $X$(3872) 
search.  The average center-of-mass energies and $e^{+}e^{-}$ integrated luminosities near $\Upsilon$($1S-5S$) and $\Lambda_{b}\bar{\Lambda}_{b}$ 
threshold are denoted by $\sqrt{s_{i}}$ and 
${\cal{L}}_{i}$($e^{+}e^{-}$), respectively.}
\begin{center}
\lineup
\begin{tabular}{lll}
\br
&$\sqrt{s_{i}}$ (GeV)&${\cal{L}}_{i}$($e^{+}e^{-}$) (fb$^{-1}$)\\
\mr
$\Upsilon(1S)$&\09.458 &1.47\\ 
$\Upsilon(2S)$&10.018&1.84\\  
$\Upsilon(3S)$&10.356&1.67\\  
$\Upsilon(4S)$&10.566&8.97\\  
$\Upsilon(5S)$&10.868&0.43\\
$\Lambda_{b}\bar{\Lambda}_{b}$ threshold&11.296&0.72\\
\br
\end{tabular}
\end{center}
\end{table}

Resonance production by untagged $\gamgam$ fusion and by ISR has similar 
characteristics. The undetected electrons in untagged $\gamgam$ fusion and 
the undetected radiated photons in ISR have angular distributions sharply 
peaked along the beam axis. Both processes have total observed energy 
(E$_{tot}$) much smaller than the center-of-mass energy, $\sqrt{s}$, of the 
original $e^{+}e^{-}$ system and have small observed transverse momentum.  
The detailed characteristics for $\gamgam$ fusion and ISR mediated $X$(3872) 
production are studied by generating signal Monte Carlo (MC) samples, using 
the formalism of Budnev {\itshape{et al.}} \cite{ggcs} for $\gamgam$ fusion 
and the formalism of M. Benayoun {\itshape{et al.}} \cite{isrprod} for ISR.

\begin{figure}[h]
\includegraphics[width=18pc]{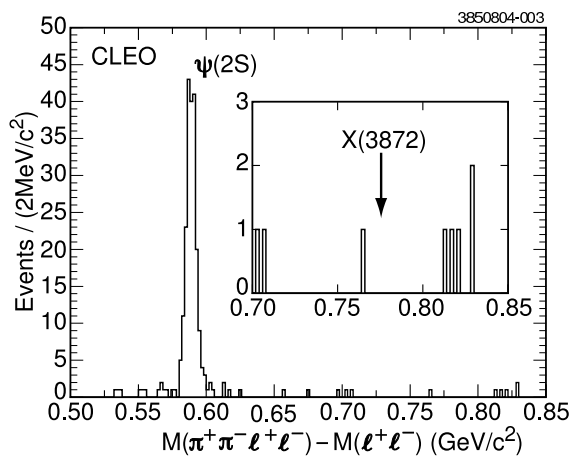}\hspace{2pc}
\begin{minipage}[b]{18pc}\caption{\label{fig:alldm}Data events as function of $\Delta$M $\equiv$ $\deltm$.  The $\psi$($2S$) is clearly visible and no apparent enhancement is seen in the $X$(3872) region.}
\end{minipage}
\end{figure}

A fully reconstructed event has four charged particles and zero net 
charge.  All charged particles are required to individually lie within the 
drift chamber volume, satisfy standard requirements for track quality and 
distance of closest approach to the interaction point, and satisfy their 
respective particle identification criteria.  Events must also have 
detected E$_{tot}$ $<$ 6 GeV, 
total neutral energy (E$_{neu}$) $<$ 0.4 GeV and total transverse momentum 
(p$_{tr}$) $<$ 0.3 GeV/c.  The lepton pair invariant mass must be consistent 
with a J/$\psi$ decay; M($e^{+}e^{-}$) = 2.96-3.125 GeV/c$^{2}$ for events 
with a $\jee$ decay and M($\mu^{+}\mu^{-}$) = 3.05-3.125 GeV/c$^{2}$ for 
events with a $\jmm$ decay.  Figure \ref{fig:alldm} shows the $\Delta$M 
$\equiv$ $\deltm$ distribution for data events which pass the selection 
criteria.  The $\psi$($2S$) is clearly visible while no enhancement is 
apparent for $X$(3872), i.e., at $\Delta$M = 0.775 GeV/c$^{2}$, which is 
indicted by the arrow in Figure \ref{fig:alldm}.

\begin{figure}[h]
\begin{center}
\includegraphics[width=22pc]{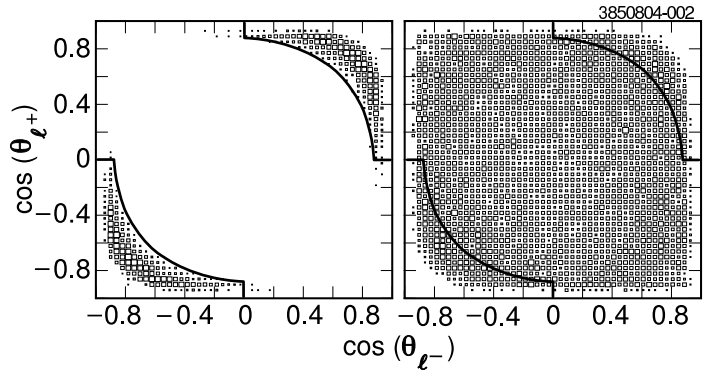}
\end{center}
\caption{\label{fig:2dcsth}MC predictions for the two-dimensional cos($\theta$) distributions for the lepton pair for ISR mediated (left) and $\gamgam$ fusion (right) $X$(3872) production.  The lines indicate how the ISR and $\gamgam$ fusion samples are separated.}
\end{figure}

At $\sqrt{s}$ $\sim$ 10 GeV, a feature unique to the ISR mediated 
production of a vector resonance which decays via $\jpipi$, $\jll$ is the 
correlation between the cos($\theta$) of the two leptons.
Figure \ref{fig:2dcsth} shows the MC prediction for the two-dimensional 
cos($\theta$) distributions for leptons from $X$(3872) decay for the 
ISR mediated and $\gamgam$ fusion productions.  As shown in Figure 
\ref{fig:2dcsth}, a parabolic cut applied to the two-dimensional 
cos($\theta$) distribution efficiently separates the events from the 
two production processes.  With this cut, the $\gamgam$ sample 
contains $\sim$86$\%$ of the $\gamgam$ events and $<$ 0.5$\%$ of the 
ISR events, and the ISR sample contains $>$ 99.5$\%$ of the ISR 
events and $\sim$14$\%$ of the $\gamgam$ events.

\section{Results}
The number of observed $X$(3872) events 
($N_{\gamgam,ISR}(X(3872))$) is determined by maximum likelihood fits of 
the $\Delta$M data distributions using flat backgrounds and the appropriate 
detector resolution functions for the two production processes.  The 
detector resolution functions are determined by the MC simulations fitted 
with double Gaussians.  The 90$\%$ confidence upper limits on the 
observed number of $X$(3872) events in $\gamgam$ fusion and ISR 
mediated production are determined to be $N_{\gamgam,ISR}(X(3872))$ 
$<$ 2.36 for both processes.

Systematic uncertainty arises from possible biases in the detection 
efficiency and estimated background level.  These are studied by varying 
the event selection criteria described above.  Other systematic 
uncertainties are from the $e^{+}e^{-}$ luminosity measurement and $\jll$ 
branching fractions.  Adding these in quadrature, the total systematic 
uncertainties in $\gamgam$ fusion and ISR are 18.5$\%$ and 
13.2$\%$, respectively.  A conservative way to incorporate these systematic 
uncertainties is to increase the measured upper limits by these amounts.  
This leads to the 90$\%$ confidence upper limits
\begin{displaymath}
\xGGUL < 12.9~\mathrm{eV}
\end{displaymath}
for $X$(3872) having positive C parity and
\begin{displaymath}
\xISRUL < 8.3~\mathrm{eV}
\end{displaymath} 
for $X$(3872) being a vector meson with $J^{PC}$ = 1$^{--}$.

\section{Summary}
With 15.1 fb$^{-1}$ of $e^{+}e^{-}$ annihilation data taken with the CLEO III 
detector near $\sqrt{s}$ = 10 GeV, we determine 90$\%$ confidence upper 
limits for untagged $\gamgam$ fusion and ISR mediated production of $X$(3872).
If $\cal{B}$($B^{\pm}$ $\rightarrow$ $K^{\pm}X(3872)$) $\approx$ 
$\cal{B}$($B^{\pm}$ $\rightarrow$ $K^{\pm}\psi(2S))$ = 
(6.8$\pm$0.4)$\times$10$^{-4}$ \cite{2004partlist} is assumed, 
we obtain $\xBR$ $\approx$ 0.02 from both the Belle \cite{xBELLE} and 
{\slshape{B{\scriptsize{A}}B{\scriptsize{AR}}}} \cite{xBABAR} results.  
This leads to 90$\%$ confidence upper limits
\begin{displaymath}
(2J+1)\Gamma_{\gamgam}(X(3872)) < 0.65~\mathrm{keV}
\end{displaymath} 
and 
\begin{displaymath}
\Gamma_{ee}(X(3872)) < 0.42~\mathrm{keV}.
\end{displaymath} 
The (2$J$+1)$\Gamma_{\gamgam}(X(3872))$ upper limit is almost 1/4 
the corresponding values for $\chi_{c0}$ and $\chi_{c2}$, 
but it is nearly 6 times larger than the prediction for 
the 1$^{1}$D$_{2}$ state of charmonium \cite{xccbartwophot}.
The upper limit for $\Gamma_{ee}$($X$(3872)) is comparable to the measured 
electron width of $\psi(3770)$ and is about 1/2 that of 
$\psi(4040)$ \cite{new4040}.

\ack
We gratefully acknowledge the effort of the CESR staff 
in providing us with excellent luminosity and running conditions.
This work was supported by the National Science Foundation
and the U.S. Department of Energy.

\end{document}